\documentstyle[seceq,epsf,graphicx,psfig]{ptptex}

\title{General Relativistic Magnetohydrodynamic Simulations of\\
Black Hole Accretion Disks: Results and Observational Implications}

\author{Julian H. {\sc Krolik} and Shigenobu {\sc Hirose}}

\inst{Department of Physics and Astronomy, Johns Hopkins University, USA}

\recdate{ }

\abst{A selection of results from the general relativistic MHD accretion
simulations described in the previous talk are presented.  We find that
the magnetic field strength increases sharply with decreasing radius and
is also enhanced near rapidly-spinning black holes.  The greater magnetic
field strength associated with rapid black hole rotation leads to a
large outward electromagnetic angular momentum flux that substantially reduces
both the mean accretion rate and the net accreted angular momentum per unit
rest-mass.  This electromagnetic stress strongly violates the traditional
guess that the accretion stress vanishes at and inside the marginally
stable orbit.  Possible observational consequences include a constraint
on the maximum spin of black holes, enhancement to the radiative efficiency,
and concentration of fluorescent Fe K$\alpha$ to the innermost part of
the accretion disk.
}

\begin{document}

\maketitle

\section{Introduction}

     In the preceding talk, John Hawley presented an overview of studies of MHD
turbulence-driven accretion onto black holes.  After outlining the principal
physical mechanisms, he summarized how to simulate numerically such a system.
At the close of his talk, he described the main structures that are seen.  In
this talk, I will report in greater detail some of the properties of the
accretion flows observed in these simulations and indicate some of their
implications for observations.

     Ultimately, it is the magnetic field
that controls how matter accretes, so the first topic in this review will be the
distribution of field intensity, the topology of field-lines, and the way both
depend on black hole spin (\S II).  Once this basis has been established, it will be
possible to see how the magnetic field controls the accretion rate (\S III).
Accreting matter releases energy via dissipation of fluid motions and magnetic field;
the heated gas can then generate photons.  Although these simulations do not treat the
thermodynamics of accretion explicitly, they do offer strong hints about how
radiation may occur; some of these are discussed in \S IV.  We conclude this
review with a list of some of the more interesting implications of our
results (\S V).

\section{Magnetic Field Structure}

\begin{figure}
   \hbox{\hskip 3cm \centerline{\includegraphics[width=15cm]{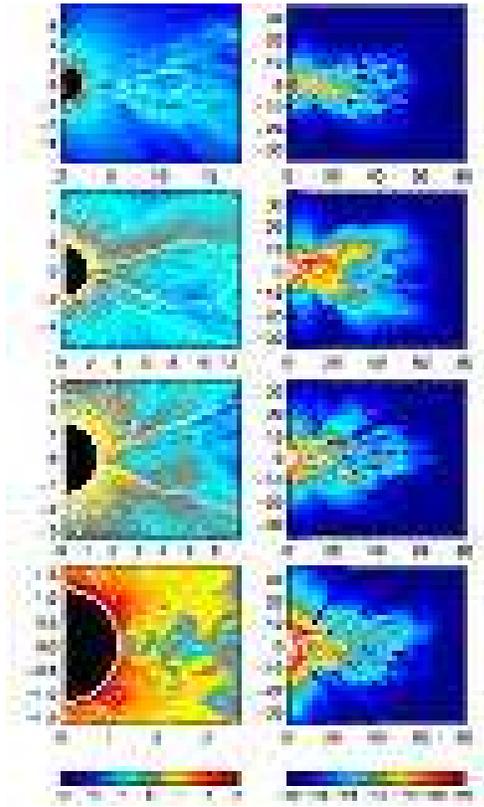}}}
   \caption{Azimuthal-average of $(1/2)||b||^2$, the magnetic pressure in
   the fluid frame, normalized to the maximum initial gas pressure in the
   relevant simulation.  The left column shows the region inside $r=3r_{ms}$,
   plotted on a logarithmic scale that emphasizes the increasing dominance of
   magnetic pressure with black hole spin in the inner disk.  The right column
   shows the main disk body with a different color scale (see color bars).  The
   white dashed contours in the right column show where the gas pressure is
   0.1, 1., and 10 times the initial maximum pressure.}
   \label{field_intense}
\end{figure}

   The most striking feature of the magnetic field is its extremely strong
concentration toward the center of the system, especially when the black
hole spins rapidly.  We can characterize the strength of the field by
the scalar $B^\mu B_\mu/(4\pi) \equiv ||b||^2$, the invariant squared-magnitude
of the magnetic field four-vector, whose value is twice the magnetic
field energy density measured in the fluid frame.  As shown in
Figure~\ref{field_intense} (and discussed more thoroughly in\cite{HKDH}),
$||b||^2$ rises steeply inward; averaged over spherical shells labelled
by Boyer-Lindquist radius $r$, the scaling is roughly $\propto r^{-3}$
all the way from the inner radial boundary to the outer, a dynamic
range of roughly two orders of magnitude in radius and six in magnetic
energy density.  Although the field intensity is more or less spherically
distributed just outside the event horizon, it becomes flattened at those
radii where the main disk body is found.  Near the main disk body, the
field is strongest near the disk surface, at the base of the corona.

    It is also useful to characterize the field strength by its ratio to
other relevant energy densities such as the gas's pressure and inertia.
The ``plasma $\beta$" is the ratio of gas to magnetic pressure.  Inside
the main disk body, we find it generally rather large, $\sim 10$--100.
However, rising vertically away from the midplane, the gas density falls
faster than the field strength, so that in the disk corona, $\beta \sim 1$,
with excursions of factors of several both up and down.  The ratio
$||b||^2/(\rho h)$, where $h$ is the specific enthalpy, indicates the
relative importance of the gas's inertia to its (magnetically-driven) dynamics.
When this ratio is $\gg 1$, the system is said to be ``force-free", and
there is a very large literature searching for ``force-free" solutions to
plasma dynamics near black holes.  We find that this criterion is satisfied
only in the outflow funnel, where the density is very low.  Throughout the
main disk body, the corona, and the inner disk, plasma inertia remains
important, although the ratio of magnetic field strength to inertia rises
to $\sim 0.1$ in the outer layers of the inflow through the plunging region.

    The radial trend of magnetic intensity is especially sharp when the black hole
spin parameter $a/M \geq 0.9$.  Although the azimuthally-averaged $||b||^2$
does not change greatly with $a/M$ at $r \simeq 25M$, at $r=5M$, it grows by a
full factor of ten as $a/M$ increases from 0 to 0.998.  This sensitivity of
the field strength in the inner disk to black hole spin is likely a by-product
of the way in which black hole spin also controls the accretion rate, as
will be explained more fully in the next section.

\begin{figure}
    \centerline{\includegraphics[width=10cm]{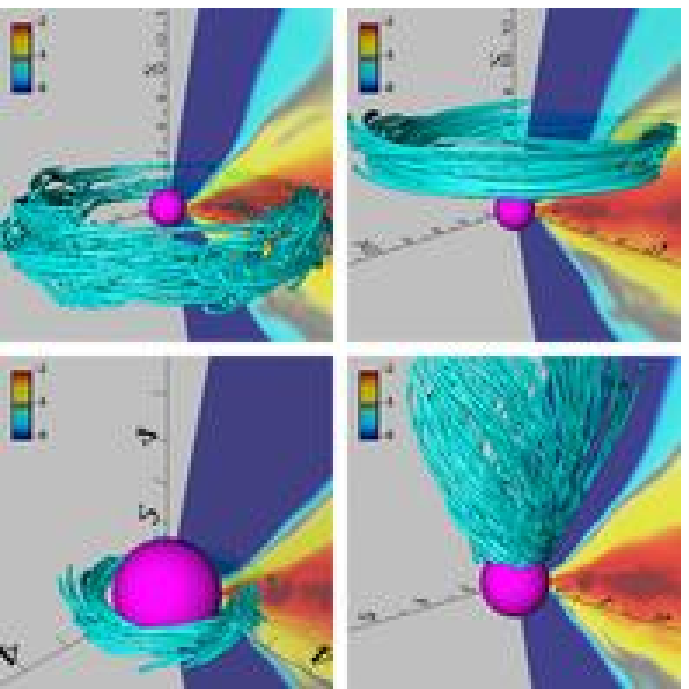}}
   \caption{Sample field-lines (as seen in the coordinate frame)
    from a late-time snapshot of the simulation with $a/M = 0.9$.  Upper left
    panel is the main disk body, upper right is the corona, lower left is
    the plunging region, lower right is the axial funnel.  Background
    colors are contours of $\log(\rho)$, the proper rest-mass density, calibrated
    by the color-bar.}
   \label{fieldlines}
\end{figure}

    In most regions, the field-line shapes are controlled by
orbital shear (Fig.~\ref{fieldlines}).  In the main disk body,
the strong turbulence creates
loops and whorls, but the shear still makes the dominant field component
toroidal.  Contrary to intuition built from the Solar corona, the field
in the disk corona does not have buoyant loops whose foot-points are twisted
by turbulent motions in the disk.  Rather, the field in the corona is
combed extremely smooth by the shear.  Because the dominant motion in
the disk is regular circular orbits, there is little opportunity for
loop-twisting.  Because the fluid trajectories in the plunging region
are dominated by ballistic infall rather than turbulence, the fieldlines
there are drawn into a pattern as smooth as the one in the corona.  The
one region where the field is notably different is the outflow funnel.
There, the field is essentally radial, as it is drawn out by the outward
flow of matter, but close to the black hole it is twisted around by
frame-dragging, as viewed in the Boyer-Lindquist coordinate frame.

\section{Magnetic Regulation of Accretion}

\begin{figure}
     \centerline{\includegraphics[width=3cm,angle=90]{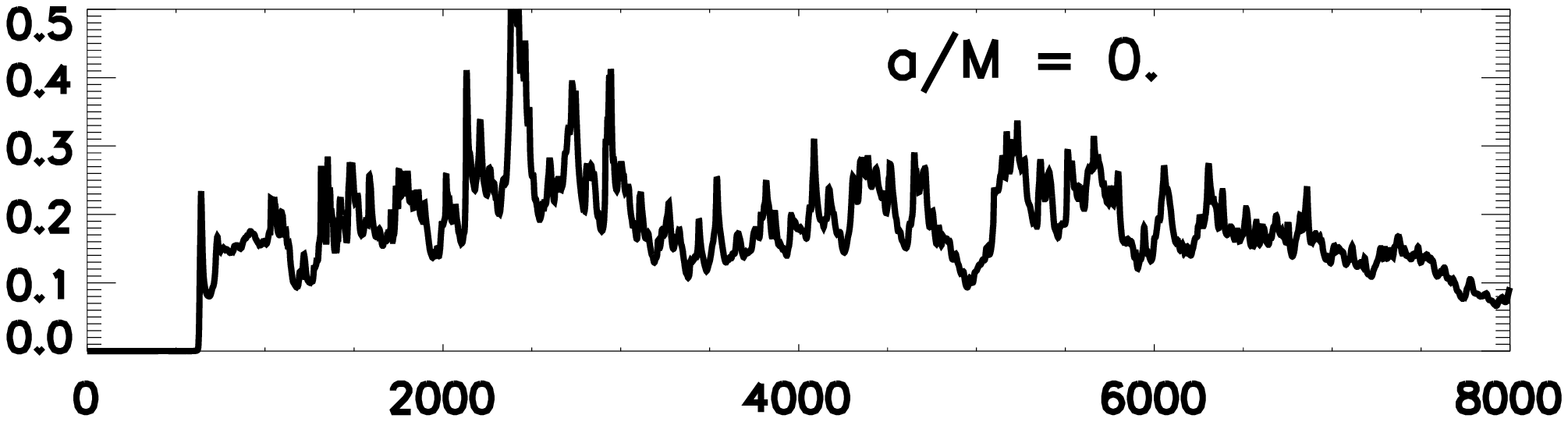}}
     \centerline{\includegraphics[width=3cm,angle=90]{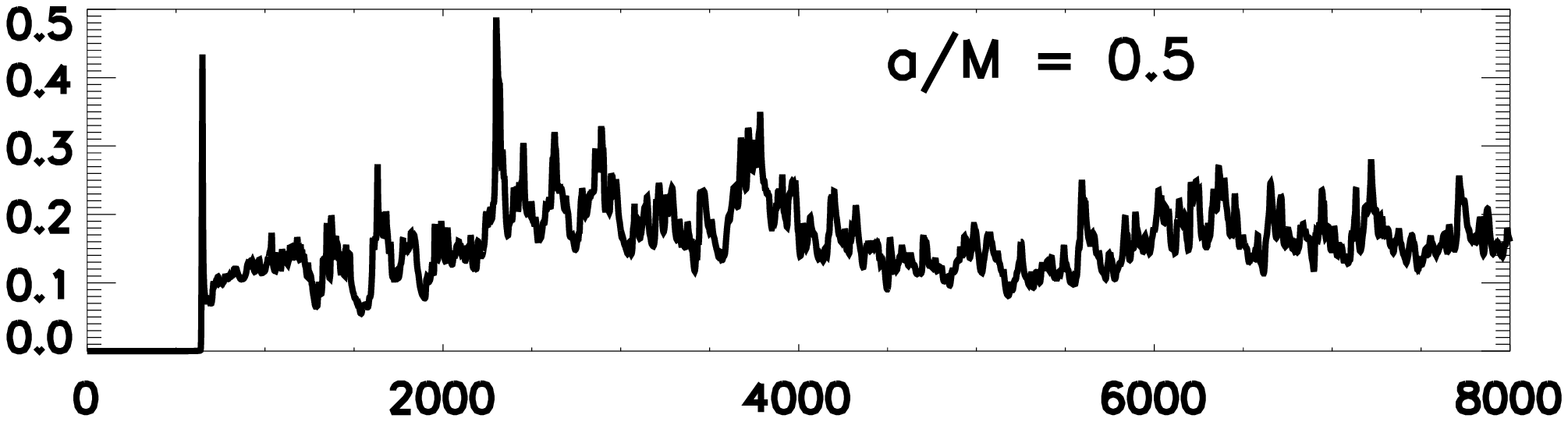}}
     \centerline{\includegraphics[width=3cm,angle=90]{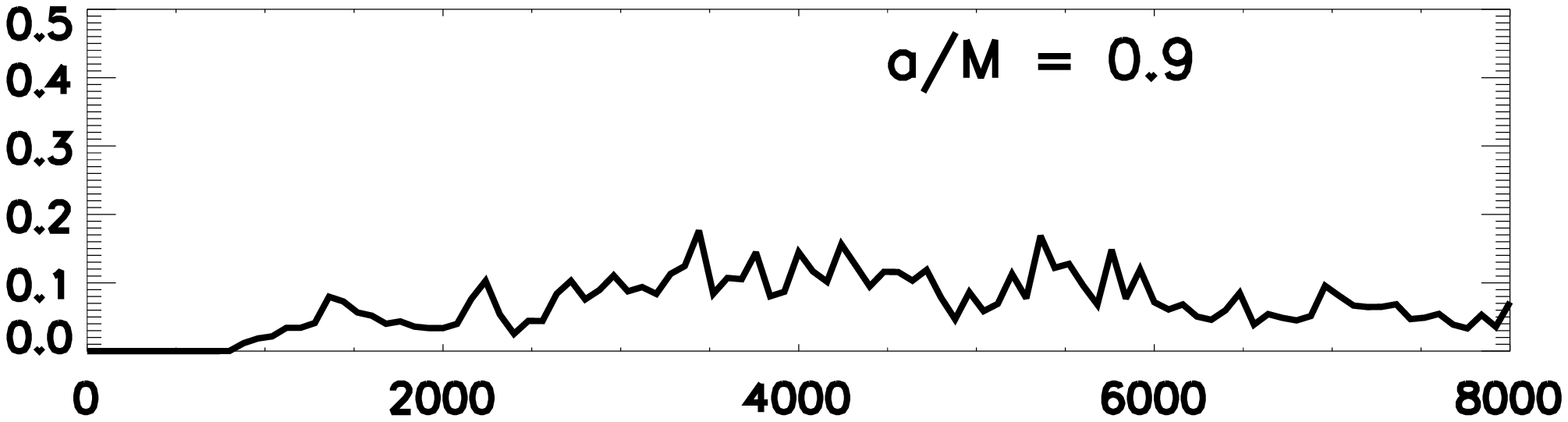}}
     \centerline{\includegraphics[width=3cm,angle=90]{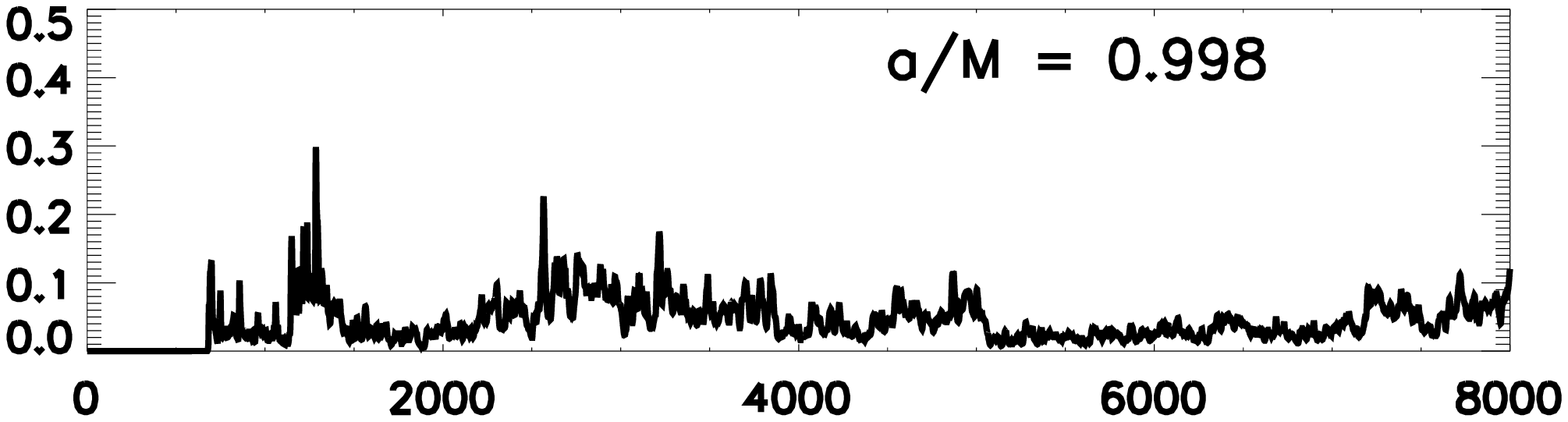}}
    \caption{The accretion rate, in units of fraction of the initial mass per
      simulation duration, for each of the four simulations.  The apparent
      absence of high-frequency variability in the $a/M=0.9$ simulation
      is an artifact of the lower data-recording rate employed for that
      simulation.}
     \label{accretionrate}
\end{figure}

     Given the central role played by turbulence in driving accretion, it should
not be surprising that the instantaneous accretion rate (i.e., the mass flux
through the inner boundary of the simulation) should vary.  What is perhaps
more surprising is two other facts (both illustrated in Fig.~\ref{accretionrate}):
that the variations occur over a wide
range of timescales, from the dynamical time at the innermost stable circular
orbit to much longer timescales; and that the mean accretion rate decreases
sharply as the black hole spin increases.

    The latter is a particularly noteworthy finding.  Although we chose
initial conditions for the four simulations that are very nearly identical,
and the black hole spin makes little difference to the potential at radii
$\gg 10M$, the amount of matter permitted to move inward through small
radii diminishes by a factor of 4 as $a/M$ rises from 0 to 0.998.  Because
the mean accretion rate just inside the initial pressure maximum (at $25M$ in
all four cases) varies hardly at all with $a/M$, the diminution in accretion
at rapid spin must---and does---result in a ``pile-up" of matter at small
radii.  In all four simulations, an ``inner torus" is created in the region
just outside $r_{ms}$, but its mass grows with $a/M$.  In fact, although
its mass reaches an equilibrium in the three simulations with $a/M \leq 0.9$,
it grows steadily throughout the $8100M$ of the simulation when $a/M = 0.998$.
That the magnetic field in the inner disk grows with $a/M$ is a reflection
of the general proportionality of the magnetic field intensity to the local
pressure in the main body of the disk, and the growth of the inner torus
with $a/M$.

     How this retardation of accretion is accomplished is most easily seen
by studying the time-average of the shell-integrated angular momentum flux.
To be more precise, the radial flux of axial angular momentum is
\begin{equation}
T^r_\phi = \rho h u^r u_\phi + ||b||^2 u^r u_\phi - b^r b_\phi ,
\end{equation}
where $u^\mu$ is the four-velocity.
It is convenient to study the three pieces separately, evaluating all quantities
in the coordinate frame.  The first we
call the angular momentum flux associated with matter, the second
the angular momentum flux of advected magnetic field, the third
the angular momentum flux of the ordinary accretion torque.

\begin{figure}[ht]
\centerline{\psfig{file=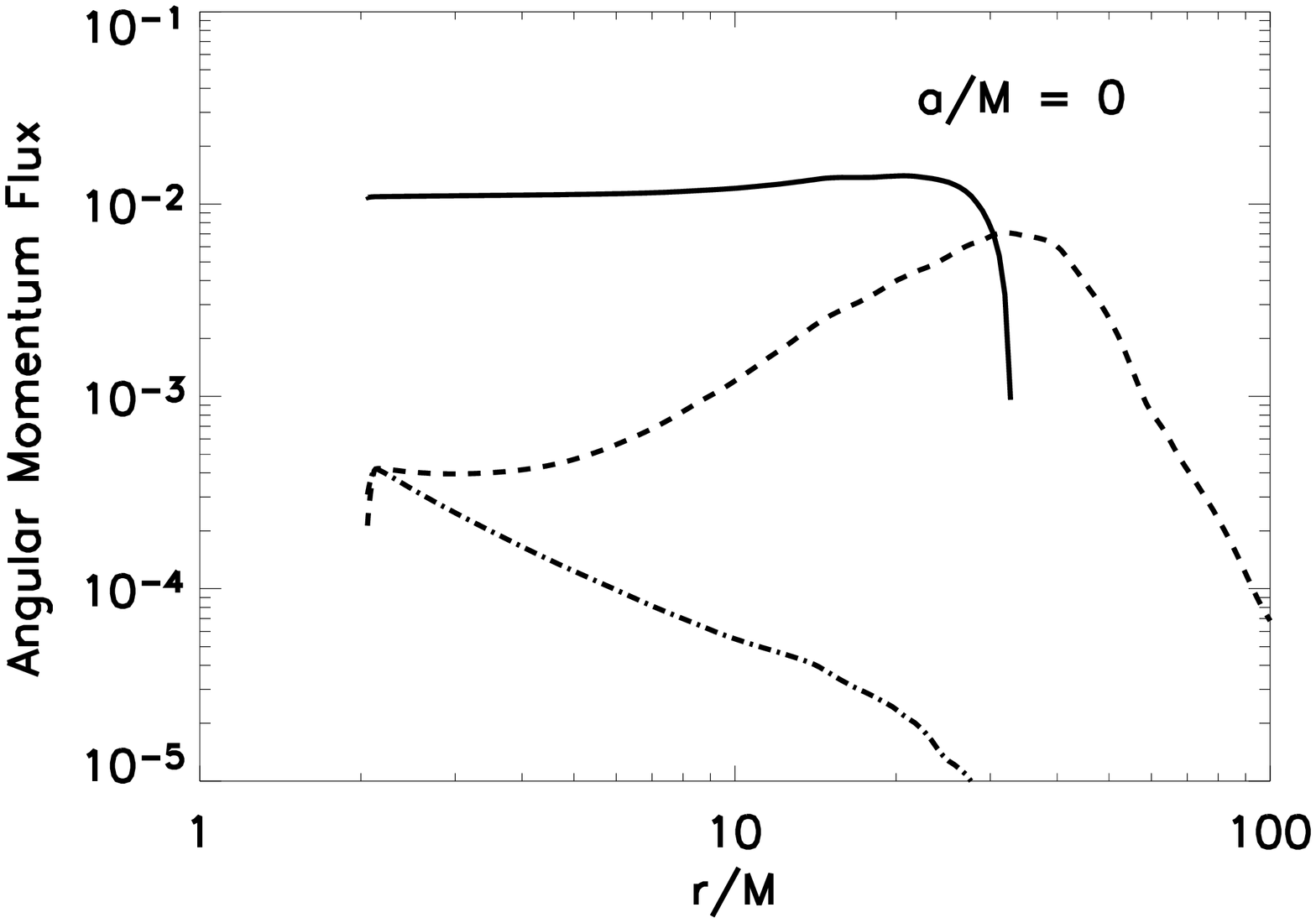,angle=90,width=2.5in}
       \quad\psfig{file=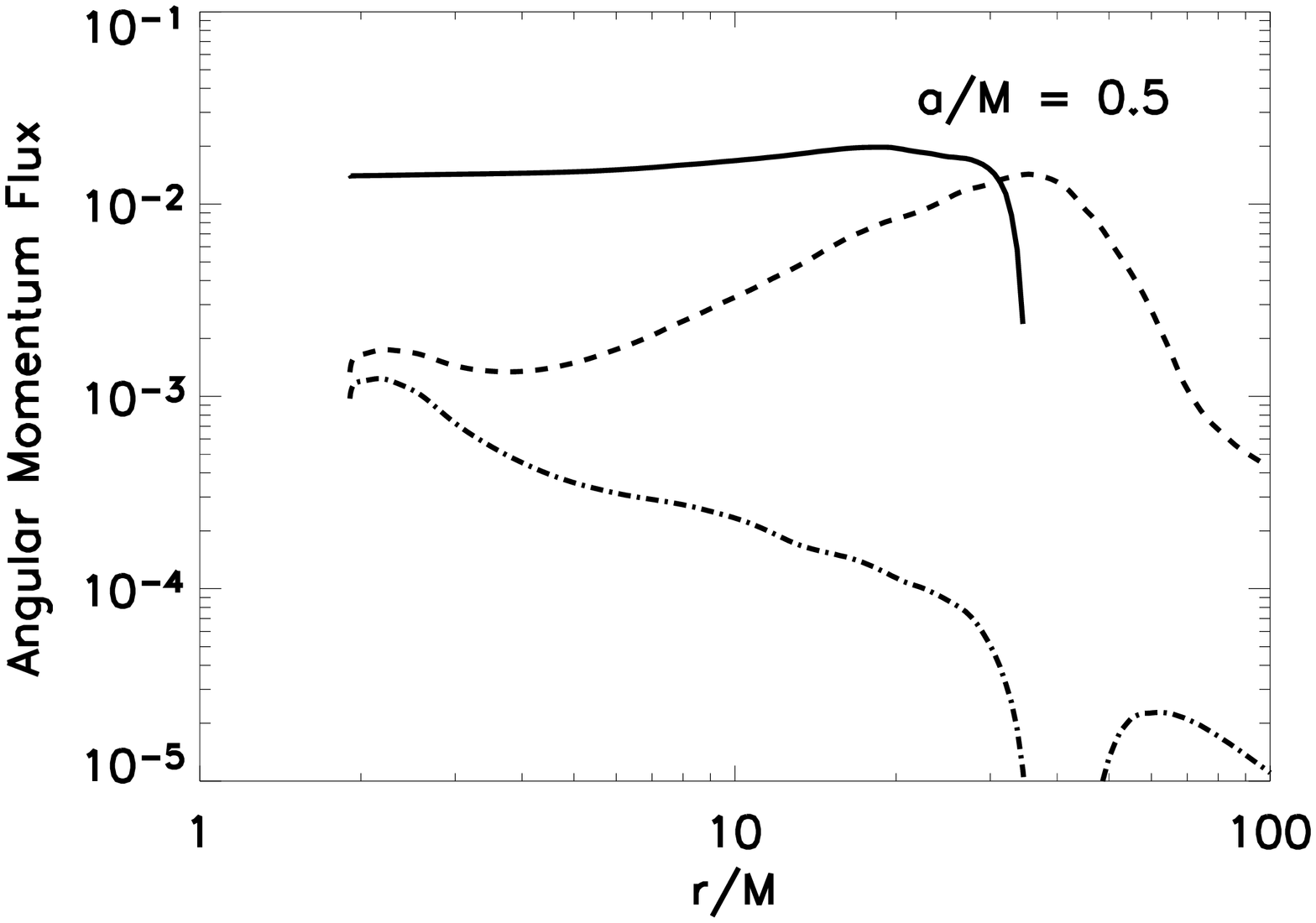,angle=90,width=2.5in}}
\centerline{\psfig{file=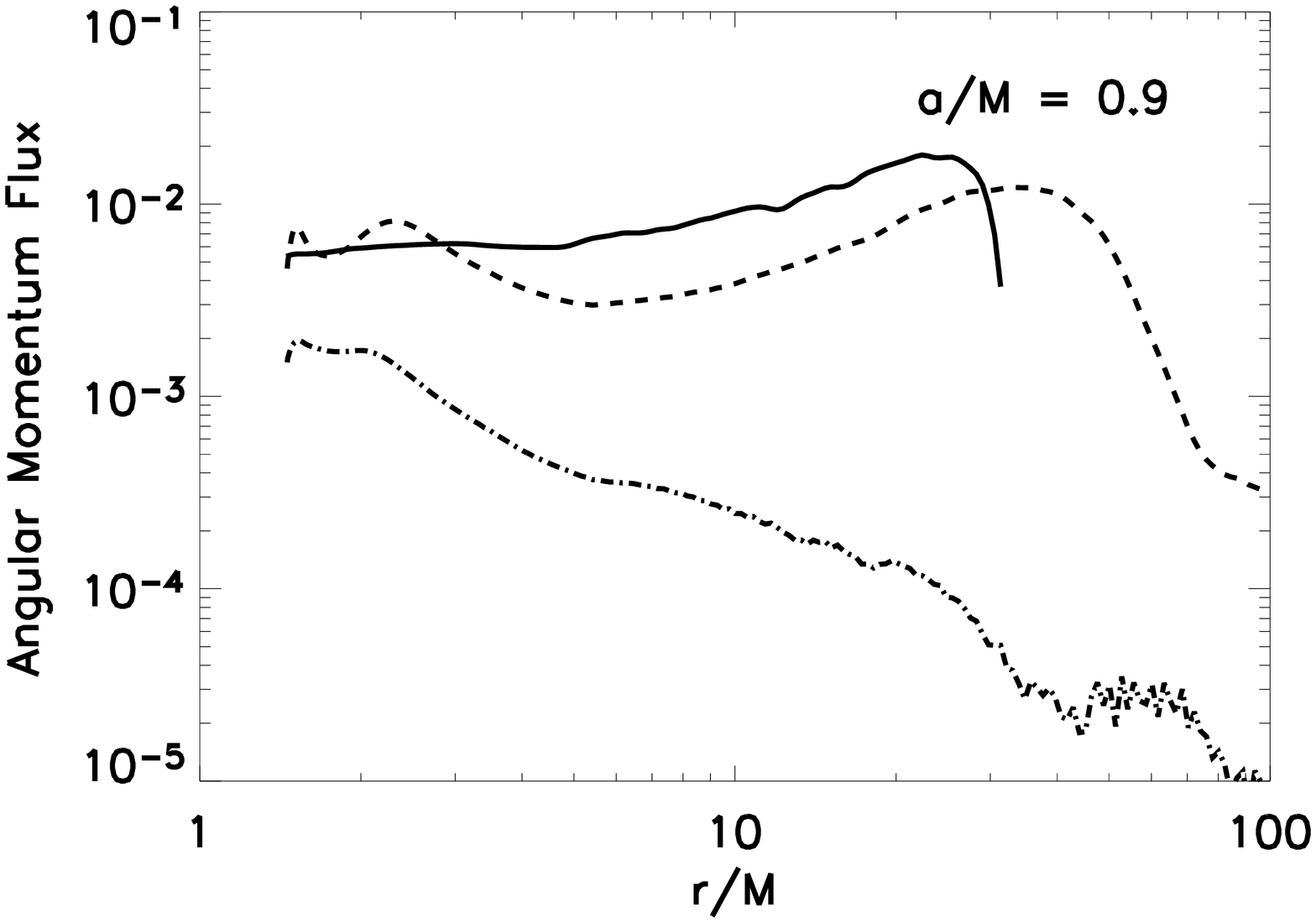,angle=90,width=2.5in}
       \quad\psfig{file=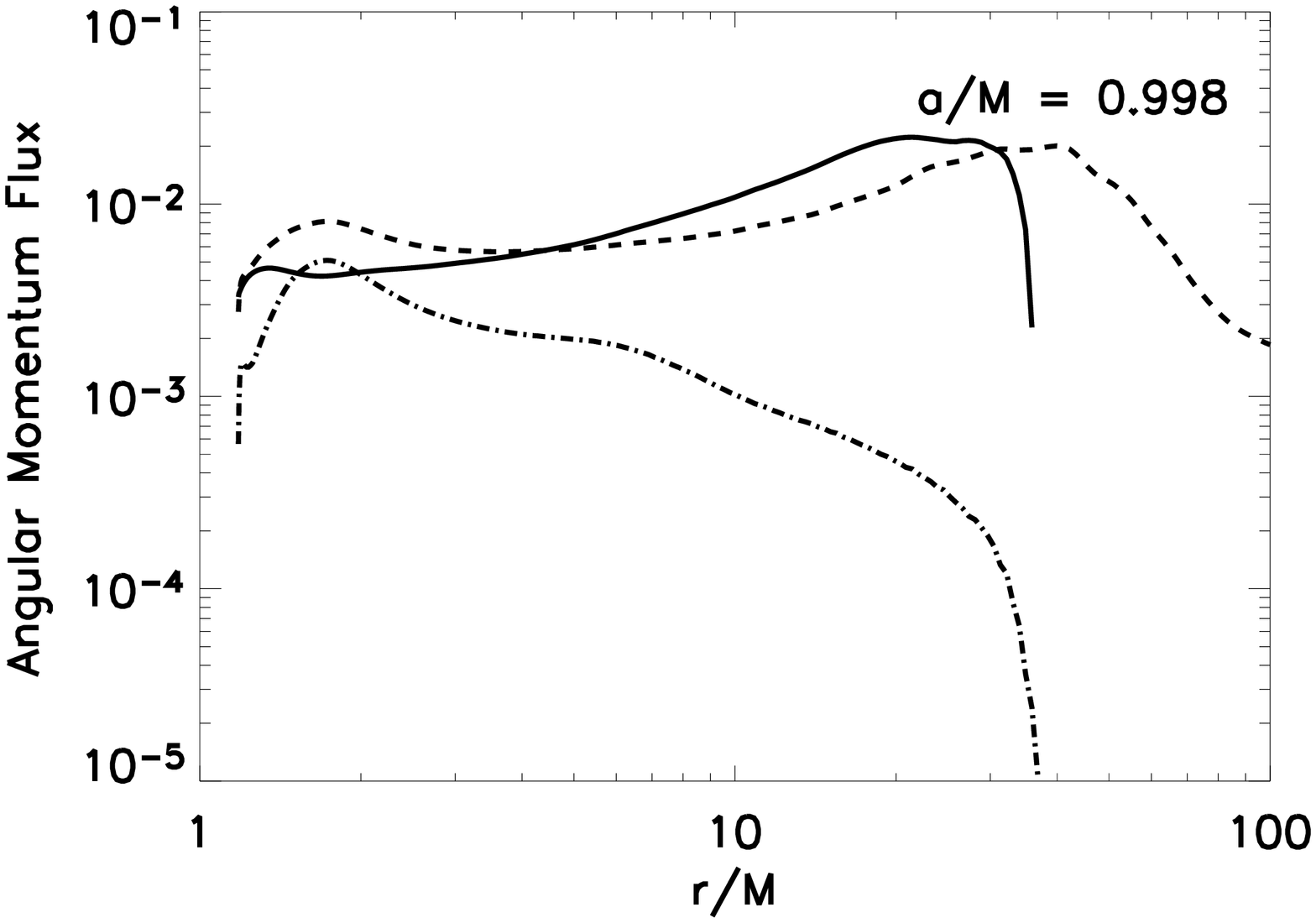,angle=90,width=2.5in}}
    \caption{\label{angmomfluxes} Absolute values of the shell-integrated and
time-averaged (for times later than $2000M$) angular momentum fluxes in each
of the four simulations.  All angular momentum fluxes are normalized to the
initial pressure maximum of the relevant simulation to ease comparison of
different simulations.  The solid curve is the matter flux, the dashed curve
is the magnetic torque, the dot-dash curve is the part due to advected
magnetic field.  The actual signs are such that the matter flux and advected
magnetic field carry (prograde) angular momentum inward, the magnetic
torque outward.}
\end{figure}

   When the magnetic torque rises outward, its net effect is to remove
angular momentum from matter, permitting the matter to move inward.  A steady
state can be achieved if those losses are balanced by gains associated with
matter moving inward.  That is the case everywhere outside $r_{ms}$ for
the $a/M = 0$ and $a/M = 0.5$ simulations, but it is {\it not} what happens
in the two simulations with more rapid spin (Fig.~\ref{angmomfluxes}).  In
both the $a/M = 0.9$ and $a/M = 0.998$ cases, the time-average magnetic
torque {\it adds} to the
matter's angular momentum over a substantial radial range starting near
$r_{ms}$ and extending outward.  Moreover, the magnitude of the outward magnetic
angular momentum flux rises sharply with increasing black hole spin,
becoming comparable to the inward angular momentum flux associated with
accreting matter when $a/M \geq 0.9$.
As a result, accretion is retarded and the mass in these rings grows.
This is the origin of both the partial suppression of accretion with
faster black hole spin and also the inner torus.

    The question immediately arises: what is the source of this outward
angular momentum flux carried by the magnetic field?  The obvious answer
is: the central black hole.  What we are witnessing here is a process
analogous to the classical Blandford-Znajek effect\cite{BZ}, in which magnetic
fields threading the event horizon of a rotating black hole are able
to carry angular momentum and energy out to infinity.  In our case,
the field lines don't precisely thread the event horizon, but that's
unnecessary: passage through the ergosphere is the essential requirement.
Likewise, the field lines of greatest importance here do not wind out
in a helix around the rotation axis, stretching to infinity; rather,
they curve around in the accreting matter, mostly staying within
$\simeq 45^{\circ}$ of the equatorial plane.

    In this context, it is important to point out that the
strength of the stresses throughout the marginally stable region demonstrates
that the heuristic ``zero-stress" boundary condition guessed thirty years
ago\cite{NT,PT} does not in fact apply.
As was feared at the time\cite{PT}, magnetic fields void the plausibility
arguments in favor of this boundary condition.

\section{Dissipation}

    In real disks, orbital shear energy is transformed into magnetic field
energy, and magnetic field energy is dissipated into heat by a variety of
resistive processes acting most effectively where there are sharp field
gradients.   Our simulations contain no explicit resistivity; instead,
purely numerical effects mimic genuine resistivity by destroying field
energy where the field gradients are large.  Because most microphysical
models of genuine dissipation predict that it rises very rapidly with
increasing current density, we can identify candidate dissipation regions
in our simulation data by computing the scalar current density-squared
$J^\mu J_\mu$, where $J^\mu = (1/4\pi)\nabla_\mu F^{\mu\nu}$ for Maxwell
field tensor $F^{\mu\nu}$ and covariant derivative $\nabla_\mu$.

    We find that the regions of strong current are highly intermittent;
that is, they tend to be organized into sheet-like structures in which
the current density can change by several orders of magnitude when
travelling a very short distance along the local normal to the sheet.
In addition, both the probability per unit volume of finding a strong
current region and the strength of the current found there increase
dramatically toward small radii.  Finally, given the substantally stronger
magnetic field of the simulations with more rapidly spinning black holes,
it is no surprise that the characteristic level of current density is
also highest when $a/M$ is relatively large.

   Because many of the most intense current regions are located within
the marginally stable orbit, we speculate that the plunging region may
be the site of substantial dissipation.  This energy release would be
over and above the conventional energy release associated with stress
in the disk body, as envisioned in the original papers\cite{NT,PT}.
If the local temperature can be raised high enough to make inverse
Compton scattering the primary radiation mechanism, the cooling time
becomes much shorter than the freefall time if there is even a modest
intensity of soft photons\cite{HKDH}.  Some of the Compton-scattered photons will
go onto capture orbits, but it is possible that the fraction escaping
could still represent an interesting addition to the radiative output
of the system.

\section{Implications}

    Our new understanding of accretion onto black holes suggests changes
in the conventional thinking about both the dynamics of accreting black
holes and their radiative properties.

    The most obvious consequence of the large diminution in the net
angular momentum per unit mass accreted is that black holes may spin up
less rapidly than thought, and it may not be possible for accretion to
make them spin faster than some maximum $a/M$, perhaps
$\simeq 0.9$\cite{GSM,DHK}.  A likely corollary of the continued
stress that reduces the accreted angular momentum is additional energy
dissipation and a consequent increase in the fraction of the accreted
rest-mass that is radiated; a quantitative determination of this
demands a more complete treatment of accretion thermodynamics and
radiation diffusion.  Another consequence of the strong outward
electromagnetic angular momentum flux, particularly when $a/M$ is
comparatively large, is the creation of a massive ``inner torus"
just outside the marginally stable orbit, a structure in the
disk surface density distribution never envisioned in the standard
treatments because they generally assumed a fixed proportionality
between stress and integrated pressure\cite{SS}.  Lastly, it is
clear from these simulations that there are consistently large
amplitude fluctuations in virtually every dynamical quantity
and across a wide range of timescales.

     Any increase in the radiative efficiency is likely to be
concentrated in the innermost part of the accretion flow.  The
extra photons produced should therefore be largely at higher
energies.  Because of the strong relativistic beaming, boosting,
and photon trajectory-bending deep in the black hole potential,
these photons will be especially noticeable when our view of
the disk is more-or-less edge-on; in addition, a significant
fraction of them will return to the disk at larger radii,
to be reprocessed there\cite{AK}.

     Additional ``coronal" activity in the inner disk
may concentrate the excitation of fluorescent Fe K$\alpha$
emission in that region.  There already
appears to be observational evidence for this\cite{W,F},
possibly including K$\alpha$ emission from the plunging
region\cite{RB,KH}.

     Finally, the ubiquitous dynamical fluctuations are a likely source
for the ``red-noise" fluctuations that are ubiquitous in black-hole
light-curves.  To predict the power spectrum of luminosity fluctuations
resulting from the dynamical fluctuations also awaits a better
treatment of accretion thermodynamics and photon diffusion, both
of which can filter the dynamical fluctuation power spectrum, but
the general scheme seems to be very promising.

\section*{Acknowledgements}
We wish to acknowledge the many contributions to this ongoing work
by our collaborators, John Hawley and Jean-Pierre De Villiers.  We also
thank Ethan Vishniac for many fruitful conversations.  This work was supported
by NSF grants AST-0205806 and AST-0313031.

\end{document}